\documentclass[a4paper,conference,onecolumn]{IEEEtran}

\usepackage{cite}      
\usepackage{graphicx}  
\usepackage{psfrag}                            
\usepackage{subfigure} 
\usepackage{url}      
\usepackage{stfloats}                          
\usepackage{amsmath}   
\usepackage{array}
\usepackage{fancyhdr}
\usepackage[colorlinks=true,dvipdfm]{hyperref}

\begin{document}

\title{A Nonlinear Map for the Decay \\ to Equilibrium of Ideal Gases}

\author{
\authorblockN{Ricardo L\'opez-Ruiz}
\authorblockA{DIIS-BIFI, Faculty of Science \\
University of Zaragoza \\ 
E-50009 Zaragoza, Spain\\
Email: 
\href{mailto:rilopez@unizar.es}
{rilopez@unizar.es}}
\and
\authorblockN{Elyas Shivanian}
\authorblockA{Dept. of Mathematics \\
Imam Khomeini Int. University \\ 
Qazvin, 34149-16818, Iran \\
Email: 
\href{mailto:shivanian@ikiu.ac.ir}
{shivanian@ikiu.ac.ir}}
}

\maketitle


\begin{abstract}
An operator that governs the discrete time evolution of the velocity 
distribution of an out-of-equilibrium ideal gas will be presented. 
This nonlinear map, which conserves the momentum and the energy of the ideal gas,
has the Maxwellian Velocity Distribution (MVD) as an asymptotic equilibrium.
Moreover, the system displays the increasing of the entropy during the decay
to the MVD.
\end{abstract}

\IEEEpeerreviewmaketitle

\section{Introduction}

A scheme inspired in economic systems that has recently been proposed \cite{lopezruiz2011}
to explain the attractivity, and then the ubiquity, of the exponential (Boltzmann-Gibbs) 
Distribution (BGD) reads as follows: 
Let $p(m) {\mathrm d}m$ denote the PDF ({\it probability density function}) of money in
a multi-agent economic system, i.e. the probability of finding an agent of the ensemble
with money between $m$ and $m + {\mathrm d}m$.
Consider now the discrete time evolution of an initial money distribution $p_0(m)$ 
at each time step $n$ under the action of an operator $\cal T$, which represents the 
average effect on the system of many random binary interactions (with number of the order 
of the system size) between pairs of agents exchanging their money.
Thus, the system evolves from time $n$ to time $n+1$ to asymptotically
reach the equilibrium wealth distribution $p_f(m)$, i.e.
$$
\lim_{n\rightarrow\infty} {\cal T}^n \left(p_0(m)\right) \rightarrow 
p_f(x)=BGD=\delta e^{-\delta x} \hskip 0.5cm with \hskip 5mm \delta=<p_0>^{-1}\,.
$$
In this case, $p_f(m)$ is the exponential distribution (BGD) with the same
average wealth $<p_f>=\delta^{-1}$ than the initial one $<p_0>$,
due to the local and total money conservation \cite{yakoven2009}. The mathematical properties
of operator ${\cal T}$ have been disclosed in Ref.\cite{lopez2011}. 
Hence, this framework not only puts in evidence that the BGD is the equilibrium distribution
if not that in this case the BGD is asymptotically reached independently of the initial wealth 
distribution given to the system, a point of view that to date was possibly lacking in 
the literature.

In this work, we extend this perspective to another problem of the same statistical nature.
Our goal is to explain the ubiquity of the MVD in ideal gases \cite{boltzmann,maxwell}. 
In the next section, we explain how to obtain an operator $T$ in the space of velocity distributions 
in order to explain the decay of any initial velocity distribution to the MVD. 
Then, the dynamical properties of this operator $T$ will be sketched and 
some examples showing its dissipative behavior are depicted. Finally, our conclusions are given.

\section{The Model. Nonlinear Map T}

Consider an ideal gas with particles of unity mass in the three-dimensional ($3D$) space. 
As long as there is not a privileged direction in the equilibrium, we can take any direction
in the space and to study the discrete time evolution of the velocity distribution in that direction.
Let us call this axis $U$. We can complete a Cartesian system with two additional orthogonal 
axis $V,W$. If $p_n(u){\mathrm d}u$ represents the probability of finding 
a particle of the gas with velocity component in the direction $U$ comprised between 
$u$ and $u + {\mathrm d}u$ at time $n$, then the probability to have at this time $n$
a particle with a $3D$ velocity $(u,v,w)$ will be $p_n(u)p_n(v)p_n(w)$. 
The particles of the gas collide between them, and after a number of interactions
of the order of system size, a new velocity distribution is attained at time $n+1$. 
Concerning the interaction of particles with the bulk
of the gas, we make two simplistic and realistic assumptions in order to obtain
the probability of having a velocity $x$ in the direction $U$ at time $n+1$:
(1) Only those particles with an energy bigger than $x^2$ at time $n$ can contribute 
to this velocity $x$ in the direction $U$, that is, all those particles whose velocities 
$(u,v,w)$ verify $u^2+v^2+w^2\ge x^2$; (2) The new velocities after collisions are equally 
distributed in their permitted ranges, that is, 
particles with velocity $(u,v,w)$ can generate maximal velocities
$\pm U_{max}=\pm\sqrt{u^2+v^2+w^2}$, then the allowed range of velocities $[-U_{max},U_{max}]$
measures $2|U_{max}|$, and the contributing probability of these particles to the velocity $x$
will be $p_n(u)p_n(v)p_n(w)/(2|U_{max}|)$. Taking all together we finally get the expression 
for the evolution operator $T$. This is: 
$$
p_{n+1}(x)=Tp_n(x) = \int\int\int_{u^2+v^2+w^2\ge x^2}\,{p_n(u)p_n(v)p_n(w)\over 2\sqrt{u^2+v^2+w^2}}
\; {\mathrm d}u{\mathrm d}v{\mathrm d}w\,.
$$

Let us remark that we have not made any supposition about the type of interactions or collisions
between the particles and, in some way, the equivalent of the Boltzmann hypothesis of {\it molecular
chaos} would be the two simplistic assumptions we have stated on the interaction of particles with
the bulk of the gas. But now a more clear and understandable framework
than those usually presented in the literature appears on the scene. 
In fact, the operator $T$ conserves in time
the energy and the null momentum of the gas. Moreover, for any initial velocity
distribution, the system tends towards its equilibrium, i.e. towards the MVD. 
This means that
$$
\lim_{n\rightarrow\infty} T^n \left(p_0(x)\right) \rightarrow p_f(x)=MVD\;(1D\;case)\,.
$$
In the presentation, we will pass review to the properties that (until today) have been 
collected here \cite{shivanian}. Other additional properties concerning the increasing of the 
entropy with time will also be given. Let us recall at this point the result that resumes the 
behavior of the system:

\vskip 0.3cm
{\bf Conjecture:} For any $p\in B$, with $B$ the set of functions whose norm $||p||$ is equal to unity,
with finite mean energy, $<x^2,p>$, and verifying $\lim_{n\rightarrow\infty} ||T^np(x)-\mu(x)||=0$,
the limit $\mu(x)$ is the fixed point 
$p_{\alpha}(x)=\sqrt{\alpha\over\pi}\,e^{-\alpha x^2}$,
with $\alpha=(2\,<x^2,p>)^{-1}$.
In physical terms, it means that for any initial velocity distribution of the gas, 
it decays to the Maxwellian distribution, which is just the fixed point of the dynamics.
Recalling that $<x^2,p>=k\tau$, with $k$ the Boltzmann constant and $\tau$ the temperature
of the gas, and introducing the mass $m$ of the particles, let us observe that the MVD is 
recovered in its $3D$ format: 
$$
MVD = p_{\alpha}(u)p_{\alpha}(v)p_{\alpha}(w)=\left(m\alpha\over\pi\right)^{3\over 2}\,
e^{-m\alpha (u^2+v^2+w^2)} \hskip 0.5cm with \hskip 5mm \alpha=(2k\tau)^{-1}.
$$

\begin{figure}[h]
\begin{center}
\psfrag{B}{} \psfrag{A}{\large\scriptsize (a)}
\includegraphics[width=6cm,height=5cm]{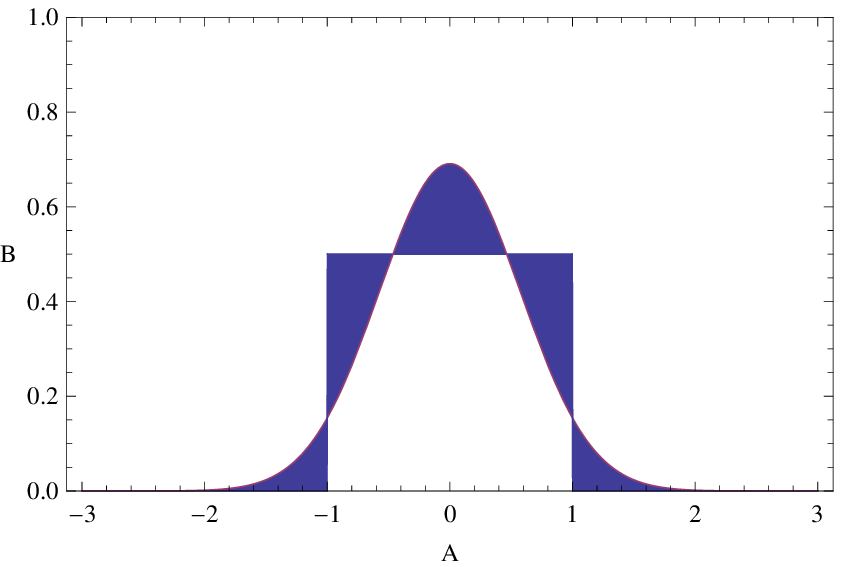} \hskip 2 mm
\psfrag{B}{} \psfrag{A}{\large\scriptsize (b)}
\includegraphics[width=6cm,height=5cm]{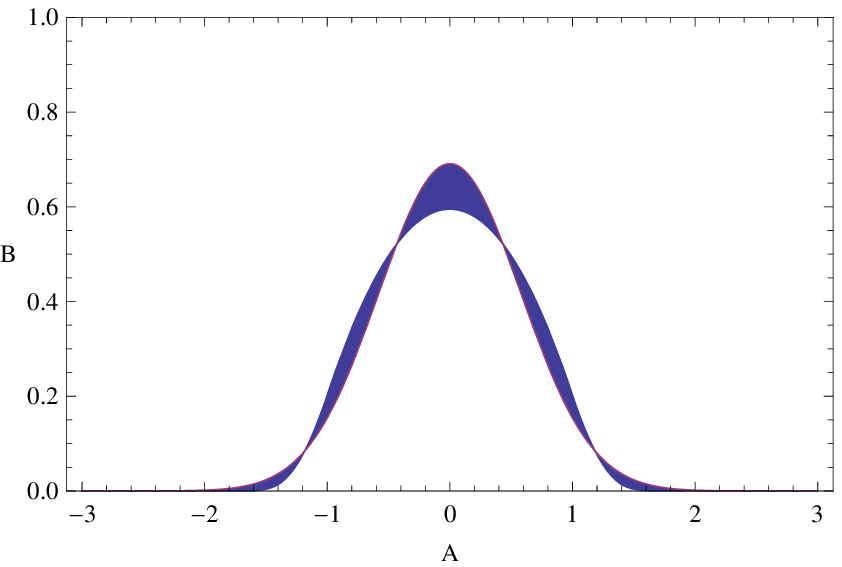} \hskip 2 mm
\caption{(a) $p(x)$ and $\mu(x)$, (b) $Tp(x)$ and $\mu(x)$.}
\label{fig-1}
\end{center}
\end{figure}

\begin{figure}[h]
\begin{center}
\psfrag{B}{} \psfrag{A}{\large\scriptsize (a)}
\includegraphics[width=6cm,height=5cm]{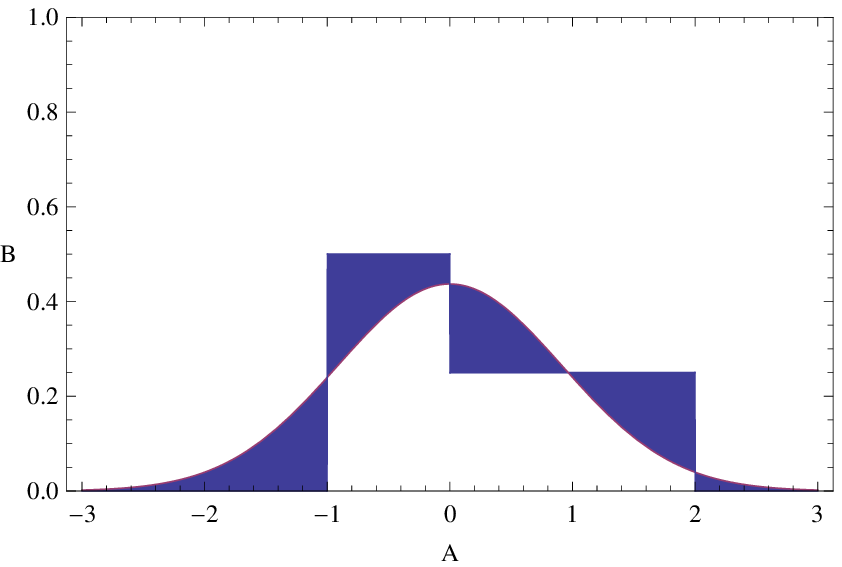} \hskip 2 mm
\psfrag{B}{} \psfrag{A}{\large\scriptsize (b)}
\includegraphics[width=6cm,height=5cm]{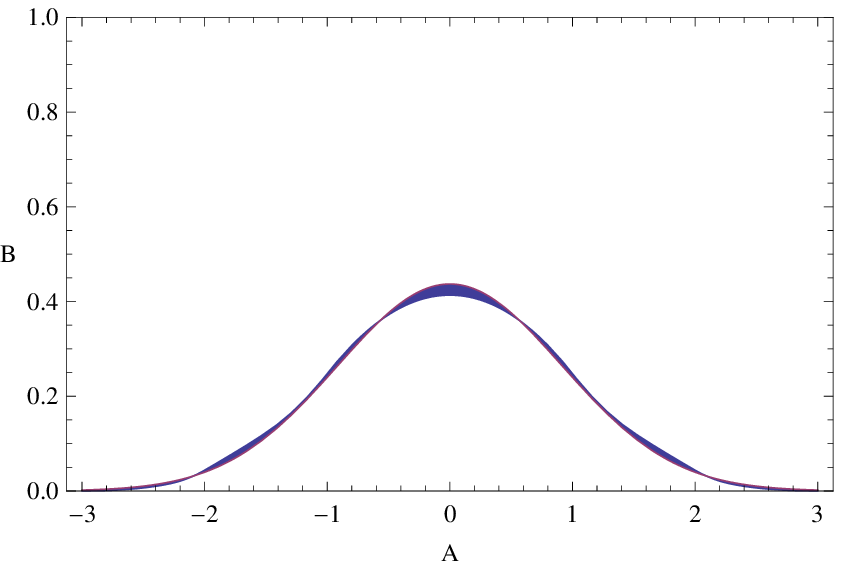} \hskip 2 mm
\caption{(a) $p(x)$ and $\mu(x)$, (b) $Tp(x)$ and $\mu(x)$.}
\label{fig-2}
\end{center}
\end{figure}

Some examples showing this behavior are depicted in the Figures \ref{fig-1}-\ref{fig-2}.

\section{Conclusions}

In this work, a nonlinear map acting on the velocity distribution space of ideal gases,
which gives account of the decay of an out-of-equilibrium velocity distribution
toward the Maxwellian distribution, has been presented. Some properties and figures
concerning its dynamical behavior have also been shown. 
  


\end{document}